\documentclass{aa}  
\usepackage[usenames,dvipsnames]{xcolor}

\usepackage{graphicx}
\usepackage{txfonts}
\usepackage{multirow}
\usepackage{pbox}
\usepackage{gensymb}
\usepackage{mathtools}
\usepackage{bm}
\usepackage{esvect}

\usepackage{xspace}
\usepackage{multicol}

\newcommand{\Msun}{\ensuremath{\,\mathrm{M_\odot}}\xspace}

\newcommand{\Msol}{\ensuremath{\,\mathrm{M_\odot}}\xspace}

\newcommand{\Rsol}{\ensuremath{\,\mathrm{R_\odot}}\xspace}

\newcommand{\kms}{\,km\,s$^{-1}$\xspace}

\newcommand{\GG}[1]{}

 \def\teff{T_{\rm eff}\,}
 
 \def\simle{\mathrel{\hbox{\rlap{\hbox{\lower4pt\hbox{$\sim$}}}\hbox{$<$}}}}
 \def\simgr{\mathrel{\hbox{\rlap{\hbox{\lower4pt\hbox{$\sim$}}}\hbox{$>$}}}}

\graphicspath{%
  {./figs/}%
  }

\usepackage{natbib,twoopt}
\usepackage[breaklinks=true]{hyperref} 

\makeatletter

\begin{document}

   \title{A model of anisotropic winds from rotating stars for evolutionary calculations}
   \author{B. Hastings  \inst{1}
		\and
            N. Langer \inst{1,2}
             \and 
			J. Puls \inst{3}	
            }
    \institute{Argelander-Institut f\"{u}r Astronomie, Universit\"{a}t Bonn, Auf dem H\"{u}gel 71, 53121 Bonn, Germany 
         \and     
    Max-Planck-Institut   f\"{u}r   Radioastronomie,   Auf   dem   H\"{u}gel   69, 53121 Bonn, Germany   
	\and 
	LMU M\"{u}nchen, Universit\"{a}tssternwarte, Scheinerstr. 1, 81679 M\"{u}nchen, Germany           
             }

    \authorrunning{Hastings et al.}

 
\abstract
   {The surface properties of rotating stars can vary from pole to equator, resulting in anisotropic stellar winds which are not included in the currently available evolutionary models.}
   {We developed a formalism to describe the mass and angular momentum loss of rotating stars which takes into account both the varying surface properties and distortion due to rotation.}
   {Adopting the mass-loss recipe for non-rotating stars, we assigned to each point on the surface of a rotating star an equivalent non-rotating star, for which the surface mass flux is given by the recipe. The global mass-loss and angular momentum loss rates are then given by integrating over the deformed stellar surface as appropriate. Evolutionary models were computed and our prescription is compared to the currently used simple mass-loss enhancement recipes for rotating stars.  }
   {{We find that mass-loss rates are largely insensitive to rotation for models not affected by the bi-stability jump.} {For those affected by the bi-stability jump, the increase in mass-loss rates with respect to time is smoothed. As our prescription considers the variation of physical conditions over the stellar surface, the region affected by the bi-stability jump is able to grow gradually instead of the whole star suddenly being affected.}   }
{We have provided an easy to implement and flexible, yet physically meaningful prescription for calculating mass and angular momentum loss rates of rotating stars in a one-dimensional stellar evolution code which compares favourably to more physically comprehensive models. 

{The implementation of our scheme in the stellar evolution code MESA is available online:} \url{https://zenodo.org/record/7437006} }


   \maketitle
%

\section{Introduction \label{sec:introduction}}

All massive stars suffer from the effects of stellar winds. For O-type stars, the winds can be so strong that a significant portion of the star evaporates and the evolutionary pathway is altered dramatically, for example forming a Wolf-Rayet star \citep{1987A&A...182..243M,1997ApJ...477..792D}. In contrast, lower-mass stars typically lose only a negligible fraction of their mass to winds. However even a non-magnetic wind carries angular momentum away from a star, and a star's spin evolution can even be affected by weak winds \citep{1998A&A...329..551L}.

In the simplest sense, {the density and velocity structure and thus also the mass-loss rate of a radiation-driven wind is determined by the opposing effects of gravity and radiative acceleration}. Gravity serves to bind material to the stellar surface, while radiation, through both continuum and line opacities, provides a force to overcome gravity \citep{1975ApJ...195..157C,1986A&A...164...86P}. Rotation directly affects both the gravitational field strength and the radiation field \citep{1924MNRAS..84..665V}, with both varying over the stellar surface, in turn resulting in an anisotropic wind \citep{1986ApJ...311..317P,1995ApJ...440..308C,2004A&A...428..545C}. 

For an anisotropic wind, attention needs to be paid to angular momentum loss since mass lost at the equator carries a larger specific angular momentum than mass lost near the poles, especially so for stars that are significantly deformed from sphericity. {Owing to internal structural changes, stars born with moderate rotation may evolve to become extremely fast rotators \citep{2020A&A...633A.165H}, so the effects of rotation on stellar winds have the potential to be relevant to a large portion of stars.}

For a number of decades, massive star modelling efforts \citep{2000ApJ...528..368H,2011A&A...530A.115B,Paxton2013} have described the effects of rotation on mass loss by increasing the mass-loss rate of an equivalent non-rotating star by a factor that depends on the rotation rate \citep{1986ApJ...311..701F}. Such a formulation is lacking due to two issues. Firstly, it is assumed that, independent of the wind recipe used, rotation always increases mass-loss rates by the same relative amount. This is not a fair assumption because two of the major effects of rotation on the surface of a star are to weaken the gravitational field and to reduce the surface-averaged effective temperature \citep{1924MNRAS..84..665V}. These two effects generally, though not always, serve to counteract each other, with winds being enhanced by weaker gravities but diminished by lower effective temperatures. It is unclear which effect dominates. Both the dependence of the mass-loss rate on temperature and the assumed temperature profile across the stellar surface (gravity darkening law) will govern whether rotation enhances or reduces mass loss, meaning that the enhancement ought to be model dependant (cf. \citealt{2014A&A...564A..57M}). 

Secondly, some of the mathematical functions used to provide the mass-loss enhancement diverge as the star approaches the critical velocity. While this behaviour is used in stellar evolution calculations merely to prevent models from exceeding critical rotation \citep{2000ApJ...528..368H,2005A&A...435.1013P}, it is unphysical not least because it is {usually\footnote{In the case of a uniform surface opacity (e.g. electron scattering opacity) and a near-Eddington luminosity, material becomes unbound over the whole surface of the star \citep{2000A&A...361..159M}.}} only material at the equator which achieves the critical rotation velocity, and strictly the equator covers an infinitesimally small surface, while gravity does manage to keep the star bound over the rest of the surface.

Angular momentum loss from massive stars plays a role in several active research topics such as the study of Be stars \citep{2004ApJ...614..929C,2005A&A...437..929C, 2008A&A...478..467E,2020A&A...633A.165H}; the occurrence of chemically homogeneous evolution, relevant to double black-hole mergers \citep{2016A&A...588A..50M} and gamma-ray burst progenitors \citep{2006A&A...460..199Y,2020ApJ...901..114A}; wind-driven orbital evolution in massive binary stars \citep{2020ApJ...902...85M,2021arXiv211103329S}; and of course the rotation rates of stars in general. Improved modelling of the winds of rotating stars would be beneficial to the advancement our understanding of stellar physics.

Various studies concerning winds from rotating massive stars have been performed (e.g. \citealt{1986ApJ...311..317P, 1989ApJ...337..888P, 1994ApJ...424..887O,1995ApJ...440..308C,2000A&A...359..695P,2000A&A...358..956P,2012ApJ...757..142C,2014A&A...564A..57M,2019A&A...625A..89G}), although the results of which have not been adopted for use in stellar evolution codes. Our aim is to provide a simple and easily implementable scheme that improves upon the popular rotationally enhanced mass-loss schemes. We applied our scheme to the one-dimensional stellar evolution code MESA and provide the files necessary to compute models using it \footnote{\url{https://zenodo.org/record/7437006}}.


The structure of this paper is as follows. Section \ref{sec:method} details the derivation of our stellar wind prescription. In Section \ref{sec:results} we compare the results of our new wind model to the commonly used rotationally enhanced mass-loss prescription. A brief discussion of uncertainties is given in Section \ref{sec:uncert}. Section \ref{sec:comp} hosts a comparison of our results to more sophisticated approaches. Lastly, our conclusions are put forward in Section \ref{sec:disc}.

\section{Method \label{sec:method}}

\subsection{Anisotropic wind model \label{sec:rot_star}}
Our basic philosophy is to apply a one-dimensional wind recipe to every point on the surface of a rotating star. For every point on the stellar surface, the given wind recipe uses the local physical conditions to provide a surface mass-flux, which when integrated results in global mass and angular momentum loss rates. We shall now determine the surface properties of a rotating star.
\subsubsection{Surface properties of a rotating star\label{sec:surface_prop}}

A rotating star with mass $M$, polar radius $R_p$, equatorial radius $R_e$, luminosity $L$ is assumed to be rotating rigidly with angular velocity $\Omega$. In reasonable agreement with detailed stellar models \citep{2009pfer.book.....M}, we assume that the polar radius is not affected by rotation. { The contribution of radiative acceleration to the total gravity shall be ignored, as we focus primarily on stars with luminosities below the Eddington luminosity. The critical velocity, or break-up velocity is then the Keplerian angular velocity, at which the gravitational force matches the centrifugal force at the equator and reads} 
\begin{align}
\Omega_{\textrm{Kep}} = \sqrt{\frac{GM}{R_e^3}} 
\end{align} 
and the fraction of Keplerian angular velocity is denoted as
\begin{align}
\omega = \frac{\Omega}{\Omega_{\textrm{Kep}}}.
\end{align}

{Recent two-dimensional models of rotating stars suggest that the rotation velocity at which material becomes unbound from the stellar surface is very close to the Keplerian velocity \citep{2019A&A...625A..88G}. However, these models only cover two separate values of the stellar mass (15 and 40\Msun) at one point in their evolution. Therefore we cannot exclude that for very luminous stars, radiation might play a significant role in unbinding material from the surface and thus reducing the critical rotation velocity. This issue is discussed further in Sec. \ref{sec:crit_vel}. }

We note that several different {working definitions of the critical velocity} velocity exist (see {discussion in Section 2.3.1 of \citealt{2013A&ARv..21...69R} and their Eqns. 3 and 4 }). Our choice is made to be consistent with the stellar evolution code MESA \citep{Paxton2019}.


In the {co-rotating} frame of a rotating star, the centrifugal force is perpendicular to the rotation axis, which causes the effective surface gravity to to become latitude-dependant. Following from the varying surface gravity , effective temperature also varies across the surface \citep{1924MNRAS..84..665V}. Also effected is the star's shape, evidenced by a bulging equator. These three effects shall now be quantified in order.

As massive stars are centrally condensed, the use of the Roche potential is justified {\citep{1965ApJ...142..265C,2016LNP...914..101R}}. We define the effective surface gravity as the sum of self-gravitation and centrifugal forces which is
\begin{align}
\vv{\bm{g}}_{\textrm{eff}} (\theta) = \left(-\frac{GM}{r(\theta)^2}\textrm{sin}(\theta) +  \Omega^2r(\theta)\textrm{sin}(\theta) \right) \vv{\bm{x}} -\frac{GM}{r(\theta)^2}\textrm{cos}(\theta) \vv{\bm{z}},
\end{align}
where $\vv{\bm{x}}$ and $\vv{\bm{z}}$ are the Cartesian unit vectors, {perpendicular and parellel to the rotation axis respectively.} The radial co-ordinate is designated $r$ and $\theta$ the co-latitude. The magnitude of the surface gravity is then found to be
\begin{align}
&|\vv{\bm{g}}_{\textrm{eff}} (\theta)| = \left(\frac{GM}{R_p^2}\right) \delta^{-2}   \Bigg[  \delta^{4}\left(\frac{r(\theta)}{R_p}\right)^{-4}  \notag \\ & +   \omega^4 \delta^{-2} \left( \frac{r(\theta)}{R_p}\right)^{2} \textrm{sin}^2(\theta) - 2\omega^2 \textrm{sin}^2(\theta) \delta \left(\frac{r(\theta)}{R_p}\right)^{-1}   \Bigg]^{\frac{1}{2}}, \label{eq:gravity}
\end{align}
where $\delta$ is the ratio of equatorial and polar radii, $\delta=\frac{R_e}{R_p}$. Because we assume that every point on the surface can be treated as an equivalent non-rotating star (i.e. we wish to reduce the two-dimensional problem of a rotating star to one dimension), and that the flux vector is nearly perfectly aligned with the gravity vector in a rotating star \citep{2011A&A...533A..43E}, the magnitude of the gravity vector is the quantity of interest, not the gravitaional field strength in the radial direction.

The local effective temperature is defined using the local flux, $\vv{\bm{F}}(\theta)$, and the Stefan-Boltzmann constant, $\sigma$ as 
\begin{align}
\teff (\theta) ^ 4 = \lvert \vv{\bm{F}}(\theta) \rvert / \sigma.
\end{align}
The effective temperature profile is given by the model of \citet{2011A&A...533A..43E}, which assumes a Roche potential and that the flux at the surface of a star is well approximated by 
\begin{align}
\vv{\bm{F}}(\theta) = -f(r, \theta)\vv{\bm{g}}_{\textrm{eff}}, \label{eq:elr}
\end{align}
which requires the energy flux to be anti-parallel to the effective gravity. This condition is fulfilled in stars with convective envelopes and is also valid to within a very fine tolerance in stars with radiative envelopes \citep{2011A&A...533A..43E}. The function $f(r, \theta)$ is found by demanding that no heat is generated in the envelope ( i.e. $\nabla \vec{F} = 0$) and reads 
\begin{align}
f(r, \theta) = \frac{L}{4\pi GM} \frac{\textrm{tan}^2 \vartheta}{\textrm{tan}^2 \theta},
\end{align}
where $\vartheta$ is the solution to 
\begin{align}
\textrm{cos}\vartheta + \textrm{ln\, tan} \frac{\vartheta}{2} = \frac{1}{3} \omega^2 \left( \frac{r}{R_e} \right) ^3 \textrm{cos}^3 \theta + \textrm{cos} \theta + \textrm{ln\, tan} \frac{\theta}{2}.
\end{align}
Alternative gravity darkening laws are available \citep{1949ApJ...110..498S,1967ZA.....65...89L, 2006ApJ...643..460L,2020ApJ...901..100L}. We note that the gravity darkening model of \citet{2011A&A...533A..43E} predicts that the equatorial flux of a critically rotating star is zero, which might be unphysical.

Lastly, the radial profile can be determined from the Roche equipotential surface (Appendix \ref{app:rot_shape}) to be 
\begin{align}
&\frac{r(\omega, \theta)}{R_p} = \notag \\ &(2 + \omega^2)  \sqrt{\frac{2+ \omega^2}{3\omega^2\rm{sin}^2(\theta)}} \textrm{cos} \left[ \frac{1}{3}\textrm{arccos}\left( \frac{3}{2+\omega^2}\sqrt{\frac{3\omega^2\rm{sin}^2(\theta)}{2+ \omega^2}} \right) + \frac{\pi }{3} \right]. \label{eq:r_surf}
\end{align}

\subsubsection{Mass and angular momentum flux}
To quantify the wind over the stellar surface, we shall use the mass-loss rate per unit surface area, or mass-flux, $\dot{m}(\theta)$, which is related to the total mass-loss rate, $\dot{\mathcal{M}}$ via 
\begin{align}
\dot{\mathcal{M}} = \int \dot{m}(\theta)dS,
\end{align}
where $dS$ represents the infinitesimal surface element. Knowledge of the star's shape allows us to compute the above integral as (cf. {\citet{2019A&A...625A..89G}})
\begin{align}
\dot{\mathcal{M}} =2\pi \int \dot{m}(\theta) r^2(\theta) \sqrt{1+ \frac{1}{r^2(\theta)}\left(\frac{\partial r}{\partial \theta} \right)^2} \textrm{sin} \theta d \theta, \label{Eq:mdot_integrate}
\end{align}
where $r(\theta)$ is given by Eq. \ref{eq:r_surf}.

The local angular momentum flux is defined as
\begin{align}
\dot{\ell}(\theta) = \dot{m} \Omega r^2(\theta) \textrm{sin}^2 \theta \label{Eq:ldot}
\end{align}
and the global angular momentum loss rate is found by integrating again over the stellar surface as  
\begin{align}
\dot{\mathcal{L}} =2\pi \int \dot{\ell}(\theta) r^2(\theta) \sqrt{1+ \frac{1}{r^2(\theta)}\left(\frac{\partial r}{\partial \theta} \right)^2} \textrm{sin} \theta d \theta. \label{Eq:ldot_integrate}
\end{align}

\subsubsection{Determining surface mass flux}

Calculating the surface mass flux of a rotating massive star requires not only knowledge of the general mechanics of radiation-driven winds but also of several rotation specific phenomena and their interplay in driving a wind. As of yet, general mass-loss recipes exist only for non-rotating stars, and even those differ significantly depending on methods and assumptions. It is felt that although the use of a non-rotating wind recipe cannot capture the fine details of physical processes in rotating stars, their use in describing rotating star winds is still beneficial and above all represents an improvement over the almost exclusively used rotationally enhanced mass-loss scheme.   

Using the effective surface gravity profile, effective temperature profile and surface shape of a rotating star, we may assign an equivalent non-rotating star
to each co-latitude of the rotating star, for which the mass-loss rate is given by a chosen recipe. This equivalent non-rotating star is defined to have the same radius, effective temperature and surface gravity as a given latitude on the rotating star. The surface mass-flux is then, in general \begin{align}
\dot{m}(\theta)  = \frac{\dot{M}\left(\lvert \vv{\bm{g}}_{\textrm{eff}} (\theta) \rvert, \teff(\theta), r(\theta), ... \right)}{4\pi r(\theta)^2}, \label{eq:mdot}
\end{align}
where $\dot{M}$ is the function provided by the non-rotating wind recipe. The only requirement for the recipe is that it is a function of,{ or can be manipulated to be a function of,} at least the effective surface gravity, effective temperature and radius.

For the calculations in this work, we shall use the mass-loss recipe of \citet{2001A&A...369..574V}, although in principle any recipe can be used. Here the mass-loss rate is a function of the stellar mass, effective temperature, luminosity and metallicity, $Z$. For our purposes, we first need to modify the input parameters of the recipe. 

{The mass and luminosity of a non-rotating star can be described using the effective surface gravity and effective temperature, provided the radius is known. This means that} at each latitude of a rotating star, an equivalent non-rotating star would have a different mass (following from the radius and surface gravity of the rotating star) and a different luminosity (following from the effective temperature and radius). To account for this, the mass-dependence must be expressed instead in terms of the surface gravity and radius. As luminosity is determined by the Stefan-Boltzmann law, an effective luminosity for each colatitude on a rotating star can be defined as 
\begin{align}
L_{\textrm{eff}}(\theta) = 4 \pi \sigma r(\theta)^2 \teff(\theta)^4.
\end{align}
This equivalent luminosity represents the luminosity that a non-rotating star, with equal surface properties of a given colatitude, would have. It is therefore this quantity that must be used in place of the luminosity in the mass-loss recipe, which becomes 
\begin{align}
\dot{m}(\theta) = \frac{\dot{M}\left( \frac{\lvert \vv{\bm{g}}_{\textrm{eff}} (\theta) \rvert  r(\theta) ^2}{G}, \teff(\theta) , L_{\textrm{eff}}(\theta) , Z \right)}{4\pi r(\theta)^2}. \label{Eq:local_mdot}
\end{align}

\subsection{Critical rotation velocity \label{sec:crit_vel}}

For a star, there exists a critical (or break-up) rotation velocity at which material becomes unbound from the stellar surface. Although a simple concept, there are several nuances which shall be discussed here. {In this work we assume that the critical velocity is attained when the centrifugal and gravitational forces balance, however in general this is likely only an approximation.}

In massive stars the force from radiation itself contributes to the force balance, and thus has been proposed to reduce the critical velocity \citep{1997ASPC..120...83L}. The acceleration produced by radiation is proportional to the flux and opacity, which are both effected by rotation. As first argued by \citet{1998A&A...339L...5G}, when a luminous star rotates very quickly, gravity darkening causes the equatorial flux to weaken strongly, suggesting that the appropriate limit is the Keplerian one. Although analysis by \citet{2000A&A...361..159M} determined that below a threshold luminosity (around 60\% of the Eddington luminosity), the radiation force indeed plays no role in unbinding material from the surface, the issue is still not clear cut{, as discussed in the following.}

Gravity darkening is traditionally described by Von Zeipel's Law, which states that effective temperature is proportional to effective gravity to the power of $\beta$, with $\beta=0.25$. Interferometric observations of rapidly rotating stars have demonstrated that gravity darkening is not as strong as predicted by Von Zeipel's Law, with lower $\beta$ values for faster rotators \citep{2007Sci...317..342M,2009ApJ...701..209Z,2011ApJ...732...68C,2014A&A...569A..10D}. These observations are generally consistent with two-dimensional numerical models \citep{2013A&A...552A..35E} and analytic gravity darkening models \citep{2011A&A...533A..43E}, but one star, $\beta$ Cassiopeiae, appears to exhibit much weaker gravity darkening than expected \citep{2011ApJ...732...68C}, perhaps exposing weaknesses in our understanding of gravity darkening. Weaker gravity darkening would result in a stronger radiative force at the equator, hence helping to reduce the critical velocity.

The surface opacity of a rotating star is also uncertain. \citet{2000A&A...361..159M} assumed that the region with the highest opacity would be the equator, as this is the coldest part of the surface. However, the centrifugal force also causes a decrease in the matter density at the equator, consequently decreasing the opacity. Two-dimensional numerical models of stars on the zero-age-main-sequence suggest that the effect of decreasing density dominates, thus a fast rotating star is predicted to have a lower equatorial opacity than an equivalent non-rotating star \citep{2019A&A...625A..88G}, meaning that radiative acceleration is unable to contribute to the force balance. However there may be some exceptions. Firstly stars may suffer the effects of opacity bumps caused by recombination of certain species (notably hydrogen, helium and iron; \citealt{1996ApJ...464..943I}), that could drastically alter the opacity profile over the surface of the star. {Secondly in very hot stars, where the opacity is dominated by electron scattering, the surface opacity is largely independant of temperature and thus spatially uniform. Such a case would need careful study to determine whether the break-up velocity is affected.}

Classical Be stars are fast rotators with a decretion disc, which are believed to be typically rotating at approximately 70\% of the Keplerian velocity \citep{1996MNRAS.280L..31P,2013A&ARv..21...69R,2016A&A...595A.132Z,2022MNRAS.512.3331D}, and in some cases even lower \citep{2010ApJ...722..605H,2016A&A...595A.132Z}. It may be reasonably supposed that an outflowing disc will affect the structure of its host star, such that the surface flux and opacities may be different when a disc is present, thus altering the break-up velocity. There is evidence to suggest that the threshold rotation rate for the Be phenomenon, assumed to be the true break-up velocity, varies with effective temperature \citep{2005ApJ...634..585C, 2010ApJ...722..605H}, with hotter Be stars rotating more slowly than their cool counterparts. It is well understood that hotter stars are closer to the Eddington limit, which may suggest that indeed in the hotter Be stars, radiative acceleration does play a role in unbinding material. 

\subsection{Numerical method \label{sec:num_method}}
In order to investigate the effect of our prescription on the evolution of both mass and angular momentum loss rates of rotating stars, we employ the one-dimensional detailed stellar evolution code MESA \citep{Paxton2019}, version 12115. The files required to compute models presented in this work are available online\footnote{\url{https://zenodo.org/record/7437006} }. The adopted physics is largely identical, except for the stellar winds, to that of \citet{2011A&A...530A.115B} and implemented in MESA as by \citet{2019A&A...625A.132S}. The models include internal angular momentum transport achieved by magnetic torques \citep{2002A&A...381..923S} which enforce near solid-body rotation during most of the main-sequence evolution.

We run two sets of models, one using the rotationally enhanced mass-loss scheme as it is usually implemented in MESA (named the standard scheme), where the mass-loss rates are first calculated using the recipe of \citet{2001A&A...369..574V} and then {following \citet{1986ApJ...311..701F}} \footnote{{see \citet{1999isw..book.....L} for a thorough description} } increased by a factor of  
\begin{align}
\left( 1- \Omega \sqrt{\frac{R_e^3}{GM (1-\Gamma)}}  \right) ^{-0.43}. \label{eq:mesamdot}
\end{align}

The second set uses mass-loss rates set by the method outlined in Sec. \ref{sec:rot_star} and is named the local scheme. Both sets rely on the wind mass-loss recipe of \citet{2001A&A...369..574V}. This wind recipe includes the bi-stability jump effect ({first found by \citealt{1990A&A...237..409P}}), where mass-loss rates are theorised to increase dramatically during the transition to temperatures cooler than approximately 22kK owing to the recombination of Fe $\textrm{IV}$ in the atmosphere {\citep{1999A&A...350..181V}}. The impact of the bi-stability jump on mass-loss rates is not certain, with \citet{2022arXiv220308218B} noting that 'the drastic $\dot{M}$ increase found in earlier models in this region might simply be an artefact of not being dynamically consistent around the sonic point, and not allowing properly for the feedback between radiative and velocity acceleration'. {The quantitative behaviour of models near the jump is also contested \citep{2008A&A...478..823M,2018A&A...619A..54V,2021A&A...647A..28K}}. We stress that our method is not confined to a particular mass-loss recipe and that several others could be used, for example those of \citet{1989A&A...219..205K,2019A&A...632A.126S,2022arXiv220308218B}.

For the standard scheme, stellar winds are assumed to be isotropic with the angular momentum loss $\dot{\mathcal{L}} $, given by 
\begin{align}
\dot{\mathcal{L}} = {j}_{\textrm{surf}} \dot{\mathcal{M}},
\end{align}
where ${j}_{\textrm{surf}}$ is the specific angular momentum of the distorted surface and $\dot{\mathcal{M}}$ the global mass-loss rate \citep{Paxton2019}. The local scheme computes angular momentum loss according to Eqns.\ref{Eq:ldot} and \ref{Eq:ldot_integrate}, taking into account both the anisotropic wind and surface deformation caused by rotation.


We compute models with a chemical mixture representing the Large Magellanic Cloud as in \citet{2011A&A...530A.115B}. Two initial masses of 10\Msol and 20\Msol are chosen to straddle the bi-stability jump. We run models with an initial equatorial rotation velocity of 300\kms until core hydrogen depletion, {defined as a central hydrogen mass fraction of $1 \times 10^{-4}$. The chosen rotation velocity} represents the typical value for early B-stars found in the Large Magellanic Cloud \citep{2013A&A...550A.109D} and corresponds to an initial critical rotation fraction of around 0.45 for both masses. 

It is also useful to assess numerical models with varying rotation rates at a fixed point in their evolution. To this end we run models with very small timesteps until the model has burnt 3\% by mass of its initial supply of hydrogen in the core ($X_c = 0.7169$). This is approximately the earliest point at which the model finds itself in thermal and nuclear equilibrium and hence is a good point in the star's evolution to investigate. We shall term the point when $X_c = 0.7169$ the zero-age-main-sequence. 

Our models are numerically stable until initial critical velocity fractions, $\omega$, of around 0.65, so to investigate stars with faster rotation, an extrapolation is performed. The local wind scheme requires as inputs the stellar mass, rotation rate, polar radius and luminosity .

The polar radius is assumed to be invariant to rotation, so this is known from a non-rotating model. For the luminosity, we extrapolate from the slower rotating models as described in Appendix \ref{app:extrap} up to $\omega = 0.9$. Using the four named quantities, the effective gravity and effective temperature profiles can be calculated (as outlined in Sec. \ref{sec:surface_prop}), and resultingly the surface mass-flux. Thus we may investigate the behaviour of our scheme for very fast rotating stars on the zero-age-main-sequence despite not having stellar models at these rotations. Stars born with moderate rotation may evolve to rotate at high critical velocity fractions owing to internal structural changes \citep{2020A&A...633A.165H}, so it is important to check the behaviour of our scheme at near critical rotations.

\section{Results \label{sec:results}}

\subsection{Mass loss on the zero-age-main-sequence}

\begin{figure*}
	\includegraphics[width=1.0\linewidth]{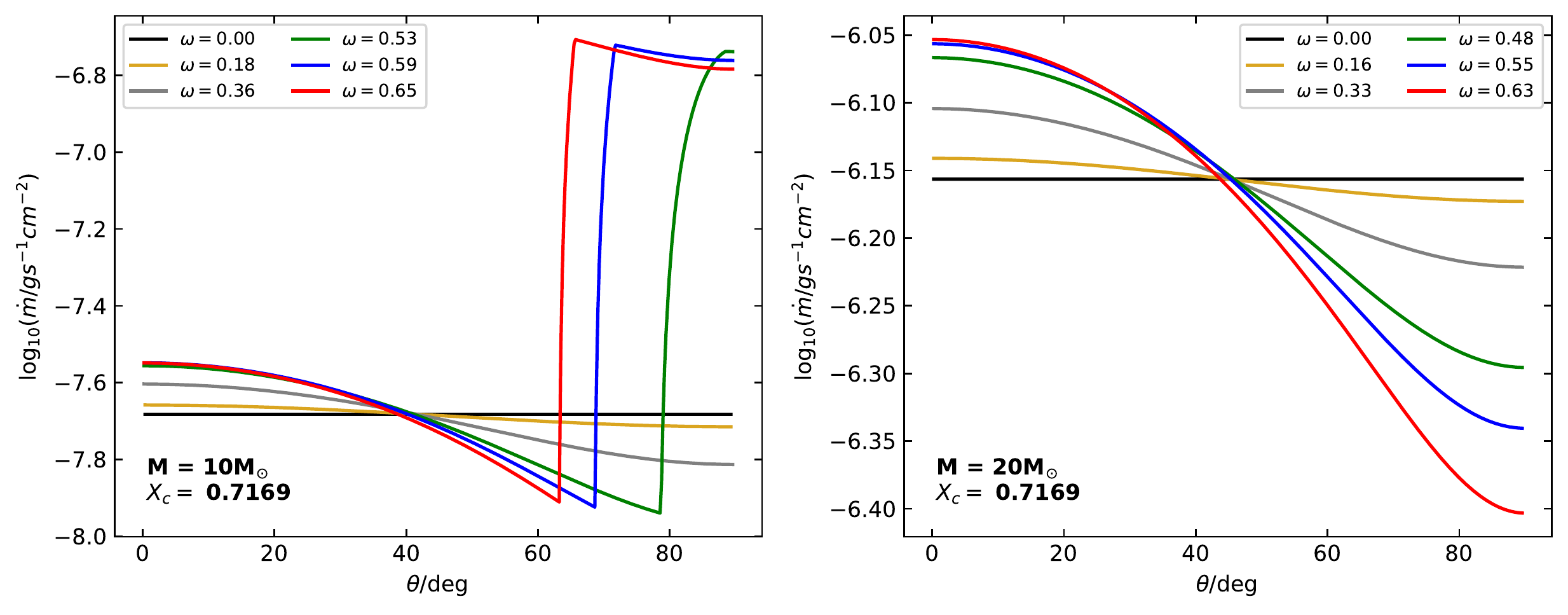}
	\centering
	\caption{Mass-loss rate per unit surface area as a function of colatitude, $\theta$, for 10\Msun (left panel) and 20\Msun models (right panel) at various rotation rates. All models have burnt 3\% by mass of their initial hydrogen (i.e. $X_c = 0.7169$). Critical velocity fraction, $\omega$, is depicted in the legend. }
	\label{fig:stationary} 
\end{figure*}

Figure \ref{fig:stationary} shows the surface mass flux as a function of colatitude for 10\Msun and 20\Msun models rotating at various rates. All models displayed have a central hydrogen mass fraction of 0.7169, equating to 97\% of the initial hydrogen mass fraction. It is seen that for slow rotation, mass flux is stronger at the poles and weaker at the equator. This occurs because rotation results in a hotter pole, relative to the non-rotating case, and a cooler equator, and stellar winds are very sensitive to effective temperature changes.

For faster rotating 10\Msun models, mass-flux experiences a jump at colatitudes between 60 and 80\degree caused by the bi-stability jump. Moving from pole to equator across the stellar surface, effective gravity and hence effective temperature decrease. At some point, the effective temperature subceeds the 'jump temperature' at which Fe $\textrm{IV}$ recombines to Fe $\textrm{III}$ causing a sudden, dramatic increase in the mass flux, as evidenced in the left panel of Fig. \ref{fig:stationary}. The 20\Msun model does not undergo the same phenomenon as here the effective temperature always exceeds the jump temperature.

\begin{figure*}
	\includegraphics[width=1.0\linewidth]{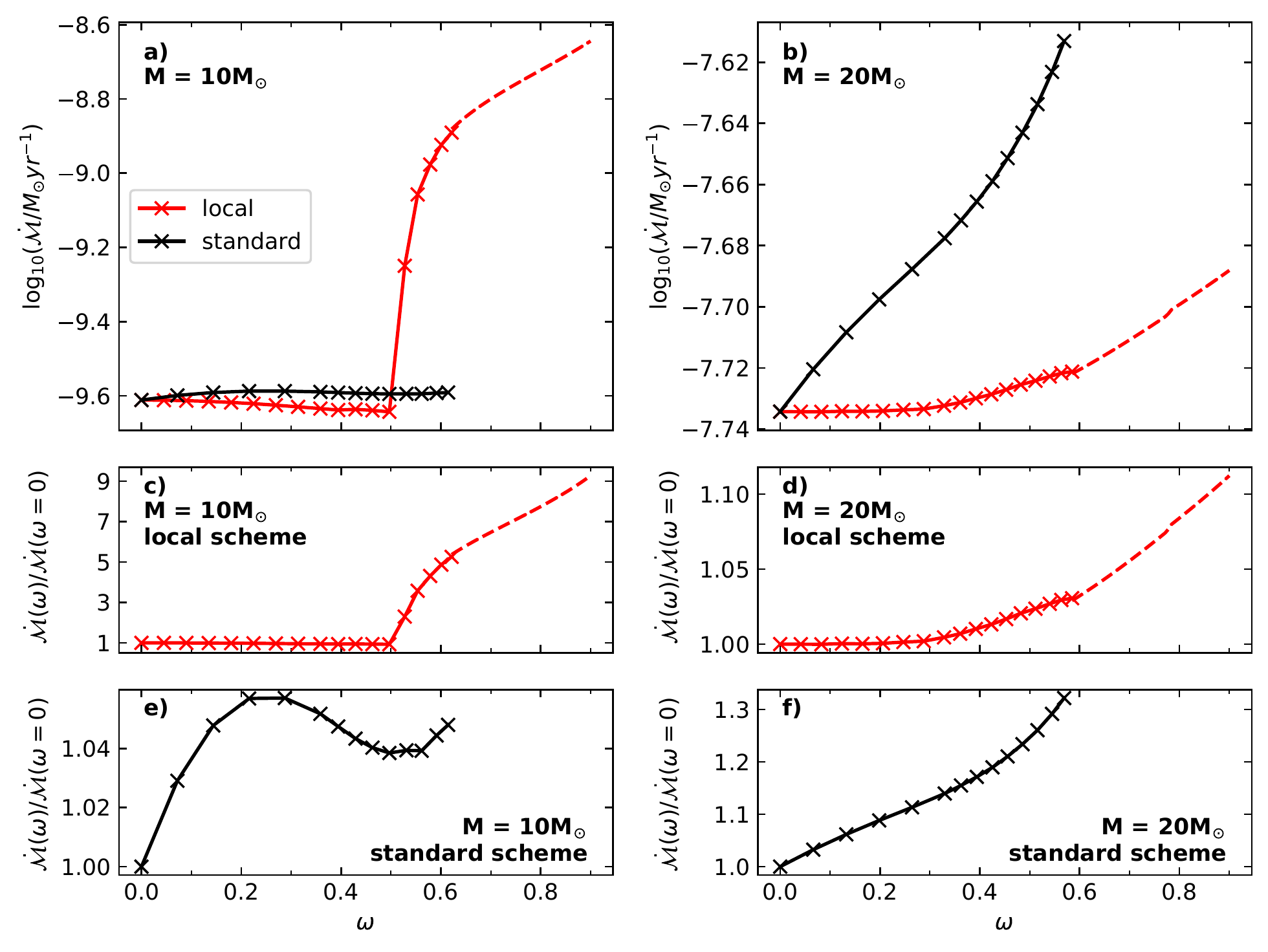}
	\centering
	\caption{\textit{Upper panels:} Global mass-loss rates as a function of critical velocity fraction, $\omega$, for 10\Msun (left panel) and 20\Msun models (right panel). {Predictions of the local scheme, where surface mass flux is determined by Eq. \ref{Eq:local_mdot} and the global mass-loss rate given by Eq. \ref{Eq:mdot_integrate}, are given in red. The standard scheme, where the mass loss rates are increased by Eq. \ref{eq:mesamdot} is depicted in black.} All models have burnt 3\% by mass of their initial hydrogen (i.e. $X_c = 0.7169$). Dotted lines represent mass-loss rates calculated from extrapolation (see Sec. \ref{sec:num_method}).
	\textit{Lower panels:} Ratios of rotating star mass-loss to non-rotating star mass loss, $\dot{\mathcal{M}}(\omega)/ \dot{\mathcal{M}}(\omega=0)$, for each combination of mass-loss model and initial mass as given in each panel.   }
	\label{fig:global_mdot} 
\end{figure*}

{The global mass-loss rate depends on both the surface mass-flux and the stellar surface area. For the wind recipe of \citet{2001A&A...369..574V} used in this work, provided the ionisation equilibrium does not change significantly, faster rotation is seen to cause a decreasing surface mass-flux at the equator. Also rotation increases the surface area of the equatorial region due to the equatorial bulge. These two effects can offset one another, causing the global mass-loss rate to be roughly independent of rotation, as exemplified by models shown in Fig. \ref{fig:global_mdot}. We note that because of the relatively small area covered by the polar region, the polar surface mass-flux does not contribute significantly to the global mass-loss rate.} 

For the 10\Msun model in the local scheme, mass loss decreases slightly with faster rotation, until the bi-stability jump comes into effect at $\omega \approx 0.5$ and drives mass-loss rate up. In contrast, the 20\Msun model displays almost no change in mass-loss rates until $\omega \approx 0.3$ and then a small increase thereafter, {due to the effect of the growing surface area of the equator dominating over the diminished equatorial surface mass-flux}. We note that, except for models affected by the bi-stability jump, our local scheme produces slightly weaker winds than the standard scheme. {Our models show that excluding the effects of the bistability jump, mass-loss rates of a rotating star are not predicted to be significantly different to that of an equivalent non-rotating star.}

{At very high initial rotation rates, our estimates of the mass-loss rate from extrapolation of the luminosity show that for the 20\Msun model, the increase in mass-loss rate is modest, 10\% at most. Whereas the cooler 10\Msun model displays mass-loss rate enhancement of a factor 9 at $\omega$ = 0.9. We are thus confident {that our scheme behaves reasonably} at near-critical velocities. }


\subsection{Evolutionary models} 

\begin{figure*}
	\includegraphics[width=1.0\linewidth]{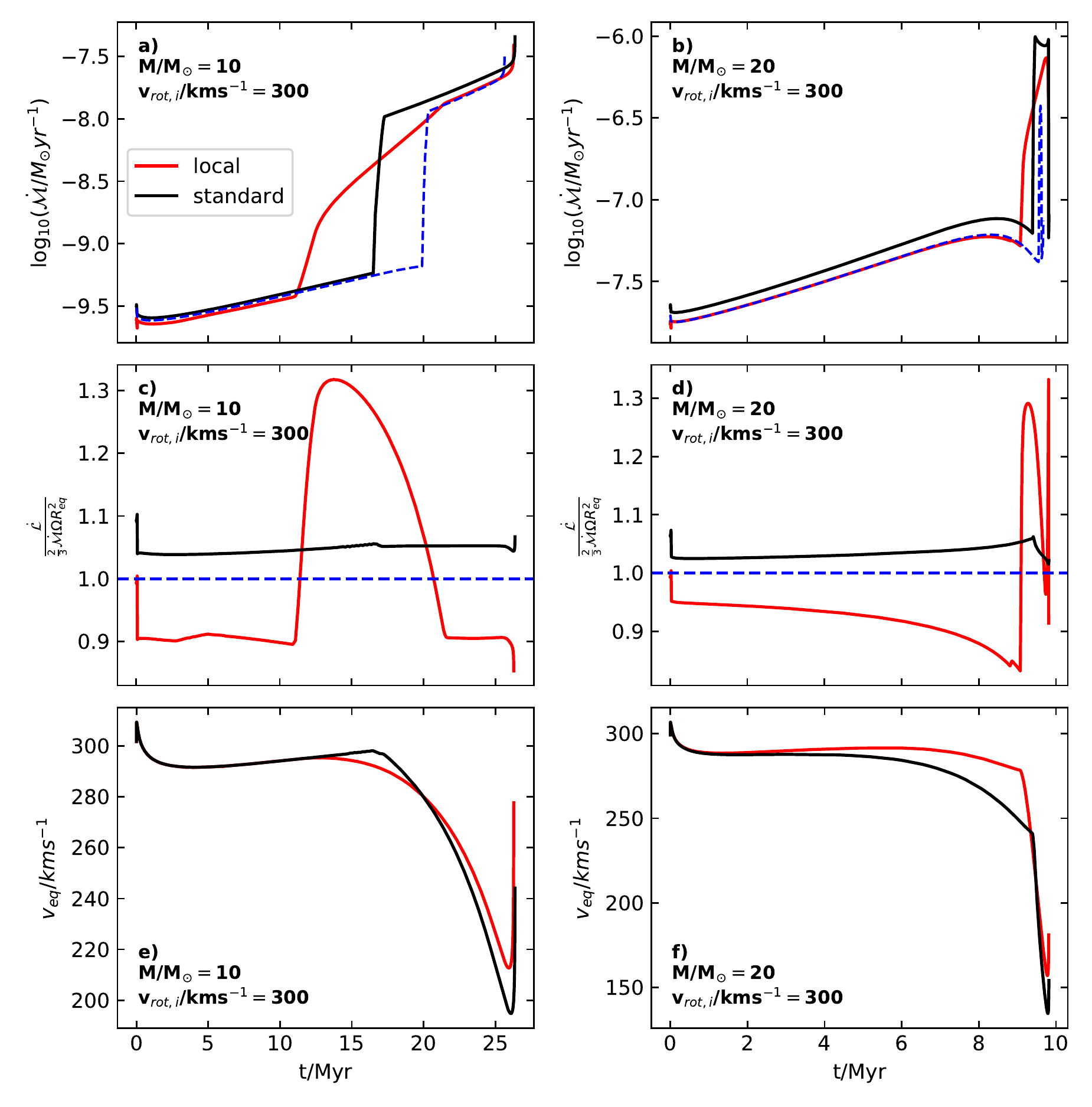}
	\centering
	\caption{\textit{Upper panels:} evolution of global mass-loss rate as a function of time. {The blue dashed line shows the mass-loss rate computed from a non-rotating model.} 
	 \textit{Central panels}: evolution of normalised specific angular momentum loss{, as given by Eq. \ref{Eq:norm_L_loss}} (see text for details). {In the limit of slow rotation} this quantity is equal to unity, shown by the blue {dashed} line.
	\textit{Lower panels:} evolution of the equatorial rotational velocity as a function of time.
The left panels show a  10\Msol model, the right panels a 20\Msol model. {Predictions of the local scheme, where surface mass flux is determined by Eq. \ref{Eq:local_mdot} and the global mass-loss rate given by Eq. \ref{Eq:mdot_integrate}, are given in red. The standard scheme, where the mass loss rates are increased by Eq. \ref{eq:mesamdot} is depicted in black.} All rotating models have initial equatorial rotational velocities of 300\kms. }
	\label{fig:MESA_mdot} 
\end{figure*}

Our evolutionary models are presented in Fig. \ref{fig:MESA_mdot} where panels a) and b) show the evolution of global mass-loss rates. In the standard scheme the bi-stability jump is implemented as a sharp jump, however in the local scheme the jump is more gradual, owing to the fact that as the star cools, the region of the surface affected by the jump grows, causing the mass-loss to also gradually increase. 


{Panels c) and d) of Fig. \ref{fig:MESA_mdot} show the normalised specific angular momentum loss of our evolutionary models, given as} 
\begin{align}
\frac{\dot{\mathcal{L}}}{\frac{2}{3} \dot{\mathcal{M}} \Omega R_{\textrm{eq}}^2}. \label{Eq:norm_L_loss}
\end{align}
This is a unitless quantity that describes the strength of angular momentum loss independently of the rotation rate and mass-loss rate. A spherical star with an isotropic wind (i.e. a slowly rotating star) has a normalised specific angular momentum loss of $1$. Values larger than unity imply that the star is losing more angular momentum per unit mass than the spherically symmetric case and that spin-down will occur more rapidly. This quantity is sensitive to both the anisotropy of the wind and the deformation of the star. We see that away from the bi-stability jump temperatures, models using the local scheme suffer lower angular momentum losses than the standard scheme. This reduced normalised specific angular momentum loss means that stars may be able to maintain faster rotation rates. The opposite is true when mass-flux across the stellar surface is increased due to the bi-stability jump, because the model loses large quantities of mass from the equatorial regions.  

Panels e) and f) of Fig. \ref{fig:MESA_mdot} show the equatorial velocities of our models. Comparing the velocities near the end of the main-sequence, we see that the local scheme displays larger rotational velocities, due to the generally lower mass-loss and normalised specific angular momentum losses as shown in the upper two panels. The effect is greatest in the 20\Msun model, with velocities increased by roughly 10\% compared to the standard scheme.


\section{Uncertainties \label{sec:uncert}}

When attempting to describe a two-dimensional {phenomenon} with a one-dimensional model there are inevitably shortcomings. Most stellar evolution codes compute the structure of a rotating star by applying certain corrections to the stellar structure equations that are designed to produce the average properties along an isobar (for a detailed description see \citealt{2000ApJ...528..368H}). This approximation may break down under certain conditions, for example when the surface temperature at the equator is cool enough for helium-I to form yet the pole it is not, the opacity will vary greatly over the stellar surface causing different physical conditions at the equator and pole. In such a case, average quantities will not capture this diversity and may lead to different structures as computed by one and two-dimensional models.

A weakness of our wind scheme is that to determine the local mass-flux, we use a mass-loss recipe that was calculated for non-rotating stars. Such a recipe naturally ignores rotational phenomena like non-spherical geometry and the effects of limb-darkening. What is more, the ionisation of the wind is expected to be sensitive to radiation from various latitudes on the stellar surface \citep{2000A&A...358..956P}, which could effect the mass-loss rates. 

A fundamental assumption of our scheme is that the wind is launched from the stellar surface and moves parallel to the photon flux (which is assumed to correspond to the direction of the effective gravity). In reality, a wind is continually accelerated until it reaches the terminal wind velocity and during this acceleration a wind particle may be influenced by photons streaming at an oblique angle to the stellar surface. This would introduce a non-radial line force (particularly in combination with a polar-angle dependent velocity field), which may alter the wind structure and angular momentum content \citep{1996ApJ...472L.115O,2000ApJ...537..461G}. {What is more, in our model the effective gravity, and hence flux, have a nonradial direction, while most one-dimensional mass-loss recipes assume a purely radial flux.}

{When running models at large critical velocity fractions, caution must be exercised, as} one may be extrapolating from the non-rotating wind recipe. For any given wind recipe, there are bounds in which the input parameters are valid and it is entirely possible that under extreme rotation the local surface conditions fall outside of the prescribed bounds. Should this occur, a second suitable wind recipe could be used to give the local mass flux for the effected regions, for example a cool star wind recipe may be appropriate for describing the equatorial wind.



\section{Applicability of the local mass-loss scheme}

{The pre-requisites for the mass-loss scheme presented here are that the star's shape needs to be well described by the Roche potential and the gravity darkening law of \citet{2011A&A...533A..43E} must be valid. This is in general true for both convective and radiative stars that do not have near-Eddington luminosities and are rotating sub-critically \citep{2011A&A...533A..43E}, however there are further cases where these conditions are not met and other special cases which will be discussed here.}

{For very luminous stars, the radiative acceleration may facilitate the unbinding of material from the stellar surface at lower rotation velocities than the Keplerian velocity. As this effect is ignored in our formalism, our scheme is not appropriate for very luminous objects. In light of the findings of \citet{2000A&A...361..159M}, we would conservatively advise the limit of applicability to be 60\% of the Eddington luminosity. Improved gravity darkening laws and a more detailed account of the stellar surface opacity could possibly allow for applying our scheme also at higher Eddington factors, but this still needs to be investigated (see Sec. \ref{sec:crit_vel}). }

{Furthermore, luminous stars may suffer the effects of inflation whereby radiation pressure "inflates" the star, producing a very tenuous, extended envelope \citep{1999PASJ...51..417I,2015A&A...580A..20S}. If the radiation pressure {deviates from spherical symmetry}, then the strength of inflation will vary according to latitude, suggesting that the star's shape is not well described by the Roche potential. Our mass-loss scheme is therefore not applicable to inflated stellar models. Inflation is expected to occur at masses above 30\Msol for stars with galactic metallicity, but for much higher masses at lower metallicities \citep{2017A&A...597A..71S}. }

{A star may suffer the effects of additional forces which can alter the surface effective gravity beyond the Roche potential. Examples include radial pulsations, accelerations from rapid expansion or contraction and a close binary companion. Such cases would need to be dealt with separately, although our scheme could be extended to them.}

{While our scheme may be applied to rapid rotators, once a star reaches critical velocity, evolutionary models demand that the star lose enough angular momentum to maintain sub-critical rotation. It is not entirely clear how this may happen, there are several possibilities. The star may undergo a "mechanical mass-loss episode", losing the required angular momentum through increased mass loss at the equator only \citep{2013A&A...553A..25G}. The other extreme is to lose angular momentum via an isotropic wind, as is currently done in MESA models, but one may also prescribe for mass to be lost from the surface in any configuration. For at least some fast rotating stars in nature, a circumstellar decretion disc forms that can efficiently drain angular momentum from the star \citep{2011A&A...527A..84K}. For lower-mass stars (M$	\lessapprox10M_{\odot}$), the required angular momentum loss rates, and correspondingly required mass loss rates are low \citep{2013A&A...553A..25G}, hence the evolution of the star is largely insensitive to the mechanics of angular momentum loss at the critical velocity. This is not true for more massive stars, which can lose upwards of 10\% of their initial mass from rotating critically (cf. Table 1 of \citealt{2013A&A...553A..25G}), so how exactly angular momentum is drained from a critical rotator becomes important. Therefore we advise caution when stellar models achieve critical rotation.} 

{A crucial aspect of our formulation is that it demands that the wind is sensitive only to the local conditions of where on the surface it was launched from. An example where this condition is violated is the dust-driven winds of asymptotic giant branch stars. Global pulsations may lead to dust formation in the outer atmosphere, which is essential for the wind driving \citep{2000A&A...361..641W}.   }

\section{Comparison to other studies \label{sec:comp}}

Several authors have investigated the problem of stellar winds and rotation by directly taking into account rotation specific physics. The prescription presented here is better described as an adaptation of a wind model for non-rotating stars, so it is useful to compare our results to previous studies.

It has been reported that radiation-driven winds are most strongly affected by gravity darkening directly beneath the point from which the wind was launched \citep{1995ApJ...440..308C}. This suggests that the wind is only sensitive to the point form which it is launched, justifying our use of a non-rotating wind model as our basis. It is also encouraging as limb-darkening, which is not accounted for in our prescription, is deemed unimportant  \citep{1995ApJ...440..308C}.

\citet{2000A&A...358..956P} calculated wind models using the concept of a mean irradiating atmosphere and found the winds to have a prolate structure, with increased mass-flux at the pole. Furthermore, for B-type stars, rotation is predicted to diminish mass-loss rates, with models rotating at around 80\% of critical velocity displaying mass-loss rates a few percent lower than corresponding non-rotating models (c.f Table 4 of \citealt{2000A&A...358..956P}). Similarly, we predict a very weak rotation dependence on mass-loss rates (away from the bi-stability jump), although our models can show enhanced or reduced winds depending on the stellar parameters. \citet{2014A&A...564A..57M} also find that mass-loss actually diminishes due to the effects of rotation, in contrast to the rotationally enhanced wind schemes.

\citet{2000A&A...359..695P} focused on B[e] stars using models including the bi-stability jump effect. They report that rotation, in general, enhances mass-flux from the poles but hardly changes that of the equator. {The spatial variation in mass-flux predicted by \citet{2000A&A...359..695P} shows a discontinuity owing to the bi-stability jump, albeit not as steep as in our results (c.f. Fig. 9 with our Fig. \ref{fig:stationary}).} \citet{2000A&A...359..695P} predict the winds of a 20\Msun star to grow stronger with rotation, with rotation at 60\% of critical velocity boosting mass-loss by 16\% compared to the non-rotating case (c.f. Table 3). 

The works mentioned above computed only stationary models, however different stellar parameters were used in each case. For example the 20\Msol model of \citet{2000A&A...358..956P} had a radius of 20\Rsol, while that of \citet{2000A&A...359..695P} was more than twice as large, 47\Rsol. The fact that the resulting relationships between rotation and mass-loss rates disagree is therefore not surprising.


{The work of \citet{2019A&A...625A..88G} differs from this study twofold. Firstly, the two-dimensional ESTER code \citep{2013A&A...552A..35E} was used to compute the stellar structure, whereas here we rely on a one-dimensional code. Secondly, the local mass-flux was calculated by calibrating the one-dimensional CAK theory \citep{1975ApJ...195..157C} to the wind recipe of \citet{2001A&A...369..574V}. Given that models presented here and in \citet{2019A&A...625A..88G} are based on the same wind recipe of \citet{2001A&A...369..574V}, a comparison between the two will highlight differences in the underlying methods.  }

\begin{figure*}
	\includegraphics[width=1.0\linewidth]{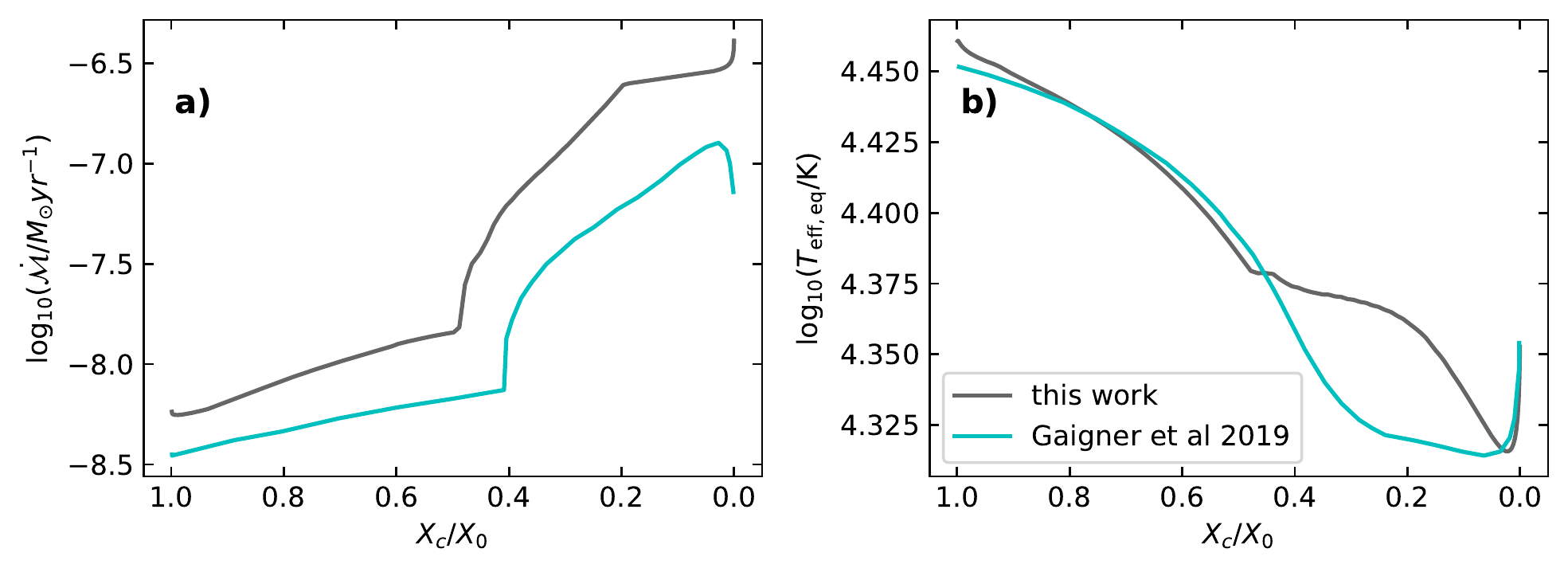}
	\centering
	\caption{Comparison of the global mass-loss rates (left panel) and equatorial effective temperatures (right panel) predicted by this work and that of \citet{2019A&A...625A..88G}. Shown are the results of 15\Msun models with solar metallicity and initial critical velocity fraction, $\omega =0.5$. Predictions of this work are plotted as grey lines, and those of \citet{2019A&A...625A..88G} as cyan lines. The $x$-axis depicts the central hydrogen mass fraction normalised to the initial value. The panels in this figure are directly comparable to Figures 13 and 14 of \citet{2019A&A...625A..88G}.    }
	\label{fig:comp} 
\end{figure*}

{To compare our scheme to that of \citet{2019A&A...625A..88G}, a 15\Msun model has been computed with solar metallicity as in \citet{2011A&A...530A.115B} and initial critical angular velocity fraction, $\omega =0.5$. Fig \ref{fig:comp} compares the results of the two methods. Firstly, we see that the equatorial effective temperatures predicted by both models agree to within around 2000K, and to within several hundred Kelvin during the early evolution. This discrepancy can largely be credited to the differences of the structures predicted by two and one-dimensional models and the implementation of rotational mixing.}

A disadvantage of the method of \citet{2019A&A...625A..88G} is that the required calibration of the surface mass-flux is sensitive to the strength of surface gravity, meaning that properly, (as stars evolve to lower surface gravities) a new calibration must be made at every timestep {(see Sec. 4.2 of \citealt{2019A&A...625A..89G}).} However, as the calibration is onerous, it was only carried out for models on the zero-age-main-sequence, meaning that "the local mass-flux may be underestimated by a factor $\sim$1.7 at most". Our scheme does not suffer from this issue, which may explain partly why our model predicts slightly higher mass-loss rates in panel a of Fig. \ref{fig:comp}. Both models however show the same general trend, with a gradual increase in mass-loss rates once the equatorial effective temperature cools below 22-23kK. The relative increase in mass-loss rates brought about by the bistability jump is approximately the same in both models. The jump temperature differs slightly in the two models because the jump temperature is sensitive to the stellar luminosity (see Eq. 16 of  \citealt{2019A&A...625A..88G}). Both models naturally have different luminosities owing to their different structures, as mentioned earlier. We find it encouraging that our relatively simple scheme behaves similarly to a two-dimensional, more advanced model.

\section{Conclusions \label{sec:disc}}

We have presented a new and simple to implement prescription for the mass and angular momentum loss rates of rotating massive stars. This represents an improvement over the widely used rotationally enhanced mass-loss schemes as we calculate the two-dimensional mass-flux over the stellar surface and are able to compute the angular momentum loss resulting from an anisotropic wind originating from a distorted star. Our method involves using a mass-loss recipe for non-rotating stars to determine the local mass-flux across the surface of a rotating star, which is then integrated to give global mass and angular momentum loss rates. 

In general we notice that, away from the bi-stability jump temperature, mass-loss rates are slightly diminished compared to the rotationally enhanced mass-loss scheme. The local mass-flux scheme has the effect of smoothing out the bi-stability jump as the increase in mass-loss rate is implemented locally on the star's surface, not globally, also observed in the models of \citet{2019A&A...625A..88G}. Our models show that the presence of the bistability-jump causes a strong relationship between rotation and mass-loss. If the bistability-jump does not infact operate in nature, as suggested by theoretical wind models of \citet{2022arXiv220308218B}, moderate and even fast rotation is not predicted to strongly alter mass-loss compared to the non-rotating case. { We see evidence that the detailed relationship between rotation and wind strength is complex, with mass-loss rates being either decreased or increased depending on the surface properties of the star.}


Our aim to provide a scheme for one-dimensional stellar evolution codes will of course mean that simplifications must be made. In spite of our scheme's shortcomings, comparisons with similar, more physically comprehensive works deliver a broad agreement in global mass-loss rates.

Our methods are relevant to several areas of stellar astrophysics where the evolution of angular momentum plays a decisive role. For example, Be stars are known to be fast rotators, so are expected to have strongly anisotropic winds and large distortions. Models such as those presented here may therefore be used to investigate evolutionary properties of Be stars. Secondly, the evolution of models along a chemically homogeneous pathway can be interrupted by spin-down caused by stellar winds \citep{2006A&A...460..199Y}. The calculations presented here suggest that angular momentum loss has been generally overestimated in stellar models, suggesting that chemically homogeneous evolution (e.g. \citealt{2020A&A...641A..86H}) may be more common or easier to maintain than previously thought. In the future, our wind prescription may be implemented in the next grids of stellar evolution models in order to gain further insights into the physics of rotating massive stars.

\bigbreak 
Acknowledgements: {The authours are grateful to Pablo Marchant for providing the MESA input files that form the basis of this work. Thanks also go to the anonymous referee who helped to improve this manuscript.} 
\bibliographystyle{aa_url}
\setlength{\bibsep}{0pt}
\bibliography{bib}{}

\clearpage


\begin{appendix}

\onecolumn
\section{Shapes of rotating stars \label{app:rot_shape}}

Here we derive the shape of the surface of a rotating star. We assume that the star is well described as a point mass enclosed by a massless envelope and that the polar radius in unaffected by rotation. The surface of the star is then an equipotential given by the Roche potential and reads

\begin{align}
\frac{GM}{r(\theta)} + \frac{1}{2} \Omega^2 r(\theta)^2 \textrm{sin}^2(\theta) = \textrm{constant}, \label{Eq:pot_shape}
\end{align}
with $M$ representing the stellar mass, $\Omega$ the angular velocity (which is assumed to be constant across both the surface and through the interior of the star), $r(\theta)$ the radial co-ordinate and $\theta$ the co-latitude. 
{We parameterise the strength of rotation with the Keplerian angular velocity, defined} using the equatorial radius $R_{\textrm{e}}$ as  
\begin{align}
\Omega^2 _{\textrm{Kep}} = \frac{GM}{R_{{e}}^3}.\label{Eq:omg_crit}
\end{align}
We let 
\begin{align}
\omega= \Omega / \Omega _{\textrm{Kep}}. \label{Eq:omg_critfrac}
\end{align}
It is important to stress that the derivation that follows is only valid for the above parameterisation of rotation.

Combining Eqns. \ref{Eq:pot_shape}, \ref{Eq:omg_crit} and \ref{Eq:omg_critfrac} gives 
\begin{align}
1 + \frac{1}{2} \omega^2 \frac{r^3}{R_{e}^3} \textrm{sin}^2(\theta) =\frac{r}{R_{e}}(1+ \frac{\omega^2}{2}), \label{Eq:a1}
\end{align}
which after defining 
\begin{align}
\tilde{r}= r /R_{{e}},
\end{align}
 further simplifies to 
\begin{align}
\tilde{r}^3 - \frac{2+ \omega^2}{\omega^2\rm{sin}^2(\theta)} \tilde{r} + \frac{2}{\omega^2\rm{sin}^2(\theta)} &=0. \label{Eq:shape_cubic}
\end{align}

Eq. \ref{Eq:shape_cubic} is a cubic in the form $x^3 + px + q =0$ (known as a depressed cubic) and has the general solution 

\begin{align}
x_k = 2 \sqrt{\frac{-p}{3}} \textrm{cos} \left[ \frac{1}{3}\textrm{arccos}\left( \frac{3q}{2p}\sqrt{\frac{-3}{p}} \right) - \frac{2\pi k}{3} \right]
\end{align}
for $k=0,1,2$ corresponding to the 3 cubic roots \citep{CRC_math_tab}.

Here we have $p= - \frac{2+ \omega^2}{\omega^2\rm{sin}^2(\theta)}$ and $q = \frac{2}{\omega^2sin^2(\theta)}$ giving 
\begin{align}
\tilde{r}_k = 2 \sqrt{\frac{2+ \omega^2}{3\omega^2\textrm{sin}^2(\theta)}}\textrm{cos} \left[ \frac{1}{3}\textrm{arccos}\left(- \frac{3}{2+\omega^2}\sqrt{\frac{3\omega^2\textrm{sin}^2(\theta)}{2+ \omega^2}} \right) - \frac{2\pi k}{3} \right]. \label{Eq:cubic_soln}
\end{align}

The periodicity of the cosine function means that if $y=\textrm{cos}(x)$ then $-y =\textrm{cos}(x + n{\pi})$ where $n$ is an integer so that $\textrm{arccos}(-y) = x +  n{\pi}$ and $x= \textrm{arccos}(y)-  n{\pi}$. Thus Eq. \ref{Eq:cubic_soln} becomes 
\begin{align}
\tilde{r}_k = 2 \sqrt{\frac{2+ \omega^2}{3\omega^2\rm{sin}^2(\theta)}} \textrm{cos} \left[ \frac{1}{3}\textrm{arccos}\left( \frac{3}{2+\omega^2}\sqrt{\frac{3\omega^2\rm{sin}^2(\theta)}{2+ \omega^2}} \right) - \frac{2\pi k}{3}-  n{\pi} \right]. \label{Eq:cubic_soln2}
\end{align}
To be physical, the solution must be independent of $n$, which is only achieved when $k=1$. {Choosing $n = -1$} results in 
\begin{align}
\tilde{r}_{k=1} = 2 \sqrt{\frac{2+ \omega^2}{3\omega^2\rm{sin}^2(\theta)}} \textrm{cos} \left[ \frac{1}{3}\textrm{arccos}\left( \frac{3}{2+\omega^2}\sqrt{\frac{3\omega^2\rm{sin}^2(\theta)}{2+ \omega^2}} \right) + \frac{\pi }{3}\right]. \label{Eq:cubic_soln3}
\end{align}

From Eqn. \ref{Eq:a1} one can deduce the ratio of equatorial to polar radii, $\frac{R_e}{R_p}$ as 
\begin{align}
 \frac{R_{{e}}}{R_{{p}}} = 1 + \frac{1}{2}\omega^2. \label{Eq:oblateness}
\end{align}

The final solution is arrived at by combining Eqns. \ref{Eq:cubic_soln3} and \ref{Eq:oblateness} and reads
\begin{align}
\frac{r(\omega, \theta)}{R_{\textrm{p}}} =(2 + \omega^2)  \sqrt{\frac{2+ \omega^2}{3\omega^2\rm{sin}^2(\theta)}} \textrm{cos} \left[ \frac{1}{3}\textrm{arccos}\left( \frac{3}{2+\omega^2}\sqrt{\frac{3\omega^2\rm{sin}^2(\theta)}{2+ \omega^2}} \right) + \frac{\pi }{3} \right]. \label{Eq:shape}
\end{align}
{As expected, the expression above gives the equatorial radius at critical rotation to be 1.5 times the polar radius ($\frac{r(\omega=1, \theta=\pi/2)}{R_{{p}}} =1.5 $). We note that Eq. \ref{Eq:shape} differs from Equation 26 used by \citet{1995ApJ...440..308C} owing to the use of different definitions of the critical velocity ({\citet{1995ApJ...440..308C} use $\Omega_{\textrm{crit}}^2 \propto {M/ (1.5R_p)^3}$ }}).

\section{Surface properties of fast rotators close to the zero-age-main-sequence \label{app:extrap}}

Our numerical models are unable to compute the structure of a rotating star with initial rotation exceeding around 70\% of the critical velocity. However the wind properties of very fast rotating stars on the zero-age-main-sequence may be determined via extrapolation. The wind scheme presented in this paper requires knowledge of the effective gravity and effective temperature profile of a star. To calculate these profiles, only the polar radius, luminosity, mass and rotation rate are required. 

\begin{figure*}
	\includegraphics[width=1.0\linewidth]{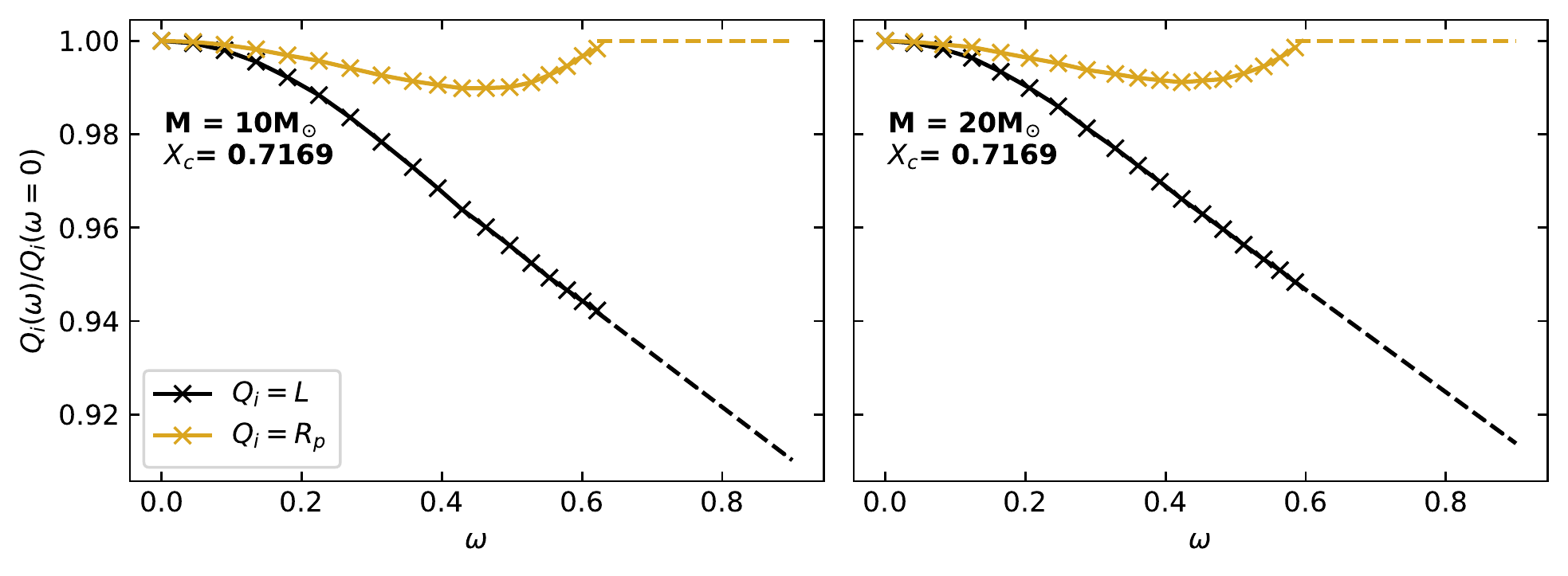}
	\centering
	\caption{Variation of luminosity (black line), $L$, and polar radius (yellow line), $R_p$, normalised to the values of a non-rotating star as a function of critical velocity fraction, $\omega$. The left panel shows models with masses 10\Msun and the right panel 20\Msun. All models have burnt 3\% by mass of their initial hydrogen (i.e. $X_c = 0.7169$). Each cross represents a value computed by a MESA model. Dashed lines show extrapolated values. }
	\label{fig:extrap} 
\end{figure*}

{To calculate the luminosity of fast rotators on the zero-age-main-sequence, we extrapolate linearly from our models with initial critical fractions between 0.4 and 0.6. We extrapolate the luminosity normalised to the value of the non-rotating model against the intial fraction of critical velocity, where all values are defined at the point when the central hydrogen mass fraction decreases by 3\% from its initial value (this we term the zero-age-main-sequence, as it is the earliest point in which the models find themselves in equilibrium). This extrapolation is shown in Fig. \ref{fig:extrap} for 10 and 20\Msun models. We find that rotation rate and luminosity are inversely proportional. This effect is rather weak, with luminosity decreasing by approximately 6\% at 60\% of critical rotation. The extrapolations suggest that at up to 90\% of the critical velocity, the luminosity is reduced by no more than 10\% compared to the non-rotating case.  }

In line with the assumptions of the Roche potential, the polar radius is not affected by rotation, so this value may be assumed from a non-rotating MESA model{, depicted graphically by the horizontal dashed yellow line in Fig. \ref{fig:extrap}.} Indeed, as evidenced by Fig. \ref{fig:extrap} the numerical models predict that to within a few percent, the polar radius remains unchanged by rotation. 

{The results of our extrapolations compare favourably to the computed stellar structures of \citet{2008A&A...478..467E}. One-dimensional models of a 20\Msun star predict that the luminosity decreases by around 8\% over the course of being spun up from stationary to critical rotation \citep[Fig. 5]{2008A&A...478..467E}. While the polar radius is judged to shrink very slightly with increasing rotation, but this is at most a 2\% effect \citep[Fig. 2]{2008A&A...478..467E}, thus justifying the assumption of the Roche potential.  }

\end{appendix}

\end{document}